\documentclass[10pt,conference,letterpaper]{IEEEtran}

\usepackage{cite}
\usepackage{amsmath,amssymb,amsfonts}
\usepackage{algorithmic}
\usepackage{graphicx}
\usepackage{textcomp}
\usepackage{booktabs}
\usepackage{multirow}
\usepackage[capitalize,noabbrev]{cleveref}
\usepackage[detect-all]{siunitx}
\usepackage{subcaption}

\def\BibTeX{{\rm B\kern-.05em{\sc i\kern-.025em b}\kern-.08em
    T\kern-.1667em\lower.7ex\hbox{E}\kern-.125emX}}

\usepackage{acro}
\newcommand{\acro}[2]{\DeclareAcronym{#1}{short=#1, long={#2}}}

\acro{CIR}{channel impulse response}
\acro{COTS}{commercial off-the-shelf}
\acro{IPS}{indoor positioning system}
\acro{LOS}{line-of-sight}
\acro{NLOS}{non-line-of-sight}
\acro{ML}{machine learning}
\acro{NN}{neural network}
\acro{RF}{radio frequency}
\acro{UWB}{Ultra-Wideband}
\acro{LR}{linear regression}
\acro{TWR}{two way ranging}
\acro{ToF}{time of flight}
\acro{MPC}{multipath component}
\acro{WAIC}{Wireless Avionics Intra-Communications}
\acro{RSSI}{received signal strength indication}
\acro{SIMD}{Single Instruction, Multiple Data}

\usepackage[numbers]{natbib}

\begin{document}

\title{Precise Onboard Aircraft Cabin\\ Localization using UWB and ML}

\author{\IEEEauthorblockN{Fabien Geyer and Dominic Schupke}%
\IEEEauthorblockA{Airbus Central R\&T, Munich, Germany}}


\maketitle

\begin{abstract}
Precise \acp{IPS} are key to perform a set of tasks more efficiently during aircraft production, operation and maintenance.
For instance, \acp{IPS} can overcome the tedious task of configuring (wireless) sensor nodes in an aircraft cabin.
Although various solutions based on technologies of established consumer goods, e.g., Bluetooth or WiFi, have been proposed and tested, the published accuracy results  fail to make these technologies relevant for practical use cases.
This stems from the challenging environments for positioning, especially in aircraft cabins, which is mainly due to the geometries, many obstacles, and highly reflective materials.
To address these issues, we propose to evaluate in this work an \ac{UWB}-based \ac{IPS} via a measurement campaign performed in a real aircraft cabin.
We first illustrate the difficulties that an \ac{IPS} faces in an aircraft cabin, by studying the signal propagation effects which were
measured.
We then investigate the ranging and localization accuracies of our \ac{IPS}.
Finally, we also introduce various methods based on \ac{ML} for correcting the ranging measurements and demonstrate that we are able to localize a node with respect to an aircraft seat with a measured likelihood of \SI{97}{\percent}.
\end{abstract}
 

\acresetall
\section{Introduction}
\label{sec:introduction}
Indoor positioning systems (\acsp{IPS}) are a necessary step towards automation across various industries.
Comprehensive research has been performed in indoor positioning using different technologies such as Bluetooth, WiFi, \ac{UWB}, each with different algorithms such as trilateration, triangulation, or fingerprinting \cite{Yassin2017,Zafari2019}.
While promising centimeter-level accuracy solutions are available when \ac{LOS} paths are available, it is hard to maintain that accuracy in harsh indoor conditions with human blockage, obstacles or reflective materials.
One of the challenging closed areas is the cabin of an aircraft, where various obstacles are present such as seats, humans, or luggage, making it significantly different from the residential, office, outdoor or industrial environments \cite{Chiu2010,Chiu2010b}.
As shown later in \cref{sec:numerical_evaluation}, this leads to an environment with mostly \ac{NLOS} conditions and with a high number of multipaths, making it generally difficult for accurate localization.

A precise \ac{IPS} would enable automatic localizing of not only wireless sensors that measure, e.g., temperature or cabin air pressure once aircraft operations started, but also other tagged items such as life vests.
In this way, cost and time savings are possible by avoiding largely manual configuration of sensors in assembly lines and by supporting cabin crew operations, such as item checks, before and after flights.

In order to be of practical use in an aircraft cabin scenario, the \ac{IPS} should be able to distinguish between cabin seats, i.e. provide an absolute localization error of approximately \SI{20}{\cm} in 2D.
\ac{UWB} was already shown to be a promising candidate technology for cabin wireless communication and for such localization requirement \cite{Karadeniz2020,Schmidt2021,Multerer2021}.
For our evaluation, we selected the Qorvo DW1000-based active localization \ac{COTS} system, based on IEEE 802.15.4 \ac{UWB}.
This solution enables both communication and localization of objects in real-time with centimeter-level accuracy.

While these previous works were limited to small mockups of aircraft cabins, we extend here our evaluation of \ac{UWB} to a larger scale measurement campaign performed in a cabin of a real Airbus A321.
Two different scenarios are evaluated, where the object to be localized is either placed near the seat or near the seat's headrest.
We perform a comparison between traditional multilateration techniques and improved methods based on \acp{NN}.
We also contribute additional methods based on \ac{NN}, and show how an end-to-end training approach achieves the best localization accuracy.
Finally, we also investigate how the number of anchors and the ranging accuracy influences the localization accuracy via Monte Carlo simulations.

Overall, our measurements illustrate the challenge that an aircraft cabin constitutes with respect to \ac{RSSI}-based and \ac{UWB}-based \acp{IPS}.
We show the influence that the cabin has on the physical properties of the \ac{RF} signals compared with outdoor and office environments free of obstacles.
We then evaluate the ranging accuracy of the system using various correction methods and show an average error of \SI{4.4}{\cm}.
Finally, we demonstrate that our \ac{IPS} is able to achieve an average localization error of \SI{16.7}{\cm} and a seat assignment error of \SI{97}{\percent}, almost fulfilling our industrial requirements for all the seats in the aircraft.

This work is organized as follows: \cref{sec:relatedwork} presents the related work and \cref{sec:localization_system} introduces the evaluated localization system and its use-cases for aicraft cabins.
\cref{sec:localization_improvements} introduces the various methods for improving the accuracy of the \ac{IPS}.
\cref{sec:numerical_evaluation} describes our measurement campaign in an Airbus A321 and localization results.
\cref{sec:discussion} gives some insights on other improvement methods via Monte Carlo simulations.
Finally, \cref{sec:conclusion} concludes this work.


\section{Related work}
\label{sec:relatedwork}

\subsection{UWB for localization}

\ac{UWB}-based localization has attracted a large body of works, due to its low cost and high accuracy.

Various works investigated the effects of the environment on the physical properties of the \ac{RF} channel.
\citet{Irahhauten2004} reported measurements and modeling of the \ac{UWB} indoor wireless channel.
They concluded that there is a limited temporal correlation between powers of multipath components and that the \ac{UWB} signal is more robust against fading than conventional narrowband and wide-band systems.

\citet{Ye2011} quantified the effect of \ac{LOS} and \ac{NLOS} on ranging in real indoor and outdoor environments.
They evaluated the impact of various materials and conditions on the ranging error and showed that \ac{UWB} is a dependable technology for ranging.
Similar studies where performed to evaluate the impact of material and conditions on ranging accuracy, such as the works from \citet{Bharadwaj2013}, \citet{Ngo2015} and \citet{Haluza2017}.

Various works investigated how to account for the \ac{NLOS} on the ranging accuracy.
\citet{Ngo2015} proposed a solution based on geometric modeling of the environment.
\citet{Bandiera2015} proposed a cognitive approach based on \ac{ML} to identify the ranging environment and estimate its relevant propagation parameters.

\citet{Schroeer2018}  used \ac{UWB}-based localization in industrial scenarios with anchors based on four transceivers instead of one to increase the system robustness.
They demonstrated an average accuracy of \SI{0.13}{\m} in situations with severe multi-path signals.

\ac{UWB} was also seen as key technology for robotics and drones, as shown by the recent survey from \citet{Wang2020}.
\citet{Hamer2018} and \citet{Tiemann2019} proposed various mechanisms for automatic configuration and calibration of a \ac{UWB}-based communication and localization network for robotic applications.

\citet{LianSang2019} proposed a novel error estimation model for \ac{TWR} and illustrate how alternative double-sided \ac{TWR} outperforms standard \ac{TWR}.
Their model includes characteristics of the conventional clock-drift error model in \ac{TWR} methods.

Alternate uses of \ac{UWB}-based ranging were also recently investigated by using the additional \ac{CIR} data, which characterizes the \ac{RF} signal propagation.
\citet{Ledergerber2019} proposed to evaluate \ac{CIR} for angle of arrival estimation.
\citet{Ledergerber2020} illustrated how \ac{CIR} post-processing can be used for building a multi-static radar network.
They demonstrated that such system can be used as passive localization by localizing a tag-free human walking in a room.


Various works also proposed a learning-based approach for correcting localization error.
\citet{Kram2019} also proposed to use the \ac{CIR} as input to a \ac{NN} in order to predict information about localization.
Similarly, \citet{Zhao2020} also used an \ac{ML}-based solution for ranging correction for small drones, but using only the ranges to the different anchors and attitude angles of the drone as input to the \ac{NN}.
Compared with these works, we propose here an end-to-end approach where the position is directly predicted, outperforming \ac{NN}-based approaches just correcting the ranges.

Finally, \ac{UWB} was recently adopted for precise localization by different phone manufacturers (e.g. Apple, Samsung) for localizing lost items or even serving as authentication mechanism.

\subsection{UWB for aicraft applications}

\ac{UWB} has already been investigated as a solution for onboard wireless communication in aircraft cabin.
\citet{Chiu2010}\cite{Chiu2010b} characterized the \ac{UWB} \ac{RF} propagation and \ac{CIR} within the passenger cabin of a Boeing 737-200 aircraft.
They reported the fading statistics and correlation properties of individual \acp{MPC}.
The effect of human presence was also evaluated and demonstrated how it affects the path gain as the density of occupancy increased from empty to full.

\citet{Andersen2012} also investigated the impact of passengers on \ac{UWB} signal absorption at the front section, upper deck of a double-decker large wide-bodied aircraft mockup.
They concluded that the absorbed power due to the presence of passengers is relatively small, so the effect of passengers was considered to be marginal for the tested configuration.

\citet{Neuhold2017} proposed a proof-of-concept for an \ac{UWB} sensor network deployed in a mockup of a small passenger cabin of a commercial aircraft with a few passengers and report experimental results on the packet loss rate.
They evaluated the loss induced by a single passenger and seat row and demonstrated packet loss rates with respect to signal attenuation.

\citet{Schmidt2021} showed how \ac{UWB} is a particularly fitting technology to support intra-aircraft communications and investigated regulation aspects.
Via a proof-of-concept implementation, they also highlighted the potential of \ac{UWB} for intra-aircraft use and identify challenges ahead.

In our previous work \cite{Karadeniz2020}, we already investigated the localization accuracy of a \ac{UWB}-based localization system, but it was limited to a small section in a cabin mockup.
We also investigated an industrial application of \ac{UWB} communication and technology for aircraft communication, more specifically with the use of \ac{WAIC} frequency range in \cite{Multerer2021}.

To the best of our knowledge, this is the first work presenting such measurement campaign of \ac{UWB}-based \ac{IPS} onboard a real aircraft cabin.
Compared also with previous works using \ac{ML} for \ac{UWB}-based localization, we are proposing a true end-to-end approach where the localization of the tag is directly predicted by the \ac{NN} based on the raw data provided during ranging.


\section{Localization system}
\label{sec:localization_system}

We introduce in this section the \ac{IPS} which was used for our evaluation.
An \ac{UWB}-based localization system was selected as a basis for our \ac{IPS} mainly because of its high accuracy, its low latency, and its strong immunity against multipath conditions compared with other solutions such as WiFi or Bluetooth-based solutions.
Those characteristics were also recognized by others, making \ac{UWB} a localization solution in mass market products such as smartphones.

\subsection{Use-cases}

An \ac{IPS} inside an aircraft cabin is highly relevant in order to assist the optimization and automation of many industrial and operational processes, from the manufacturing phase till the end-of-life phase of an aircraft.
Such system would avoid many manual and tedious tasks, leading to large time and cost savings both at aircraft manufacturing and during aircraft operation.
We review in this section various use cases within the aircraft scope and the advantages that an \ac{IPS} would bring.

It is expected to have hundreds to thousands of wireless sensors placed inside the aircraft monitoring the environment and devices status using sensors such as temperature, humidity, engine status, smoke detection, cabin pressure, seat, or door status \cite{Haowei2004,Yedavalli2011,Losada2014}.
The position of each sensor is used in order to properly correlate the measured data with the corresponding area of the aircraft.
This use-case is relevant in the final assembly line, but also during aircraft maintenance where sensors might be replaced.

Another possible application is to localize and identify seats.
Currently, the position of each seat and their corresponding seat numbers are hard-coded in a database based on the cabin configuration, dependent on each airline preference.
However, a localization system may localize seats and automatically assign seat numbers with sensors and devices placed around it.
This use-case will be evaluated later in \cref{sec:numerical_evaluation}.

Various tasks have to be performed by the cabin crew before each take-off and after landing.
Before aircraft take-off, the presence of safety equipment such as life vests, fire extinguishers, first aid kits, or portable oxygen equipment has to be checked.
An \ac{IPS} could automatically identify the location of items to be checked and give a warning if a given object is missing or not at its expected place.

In case the layout of a cabin is updated, an \ac{IPS} could assist with the reconfiguration of various devices and sensors,
automatically assigning them to a given seat position.
The new cabin layout may be automatically extracted based on localization data, and adopted for other applications by finding the new locations of sensors mounted on the seats, the floor or in the cabin luggage compartments.

Finally, regarding industrial applications and regulations, various works already demonstrated that \ac{UWB} would be a compliant technology with respect to regulations for aircraft and cabin communications \cite{Schmidt2021,Multerer2021}.

\subsection{UWB-based ranging}

Our system is based on the Qorvo EVB1000 and DWM1001 evaluation boards, a system promising centimer-level accuracy with a measurement latency of a few milliseconds.
The Qorvo platform provides an easy-to-use system for working with \ac{UWB} and supports custom software running directly on the micro-controller of the evaluations boards.

\begin{figure}[h!]
	\includegraphics[width=\columnwidth]{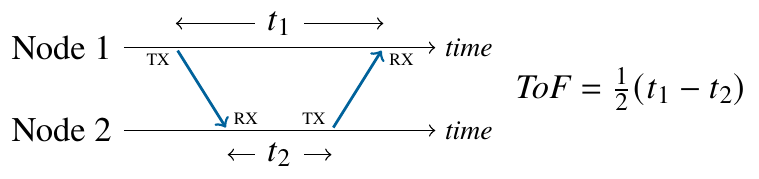}
	\caption{Illustration of single sided \acf{TWR} and \acf{ToF} calculations}
	\label{fig:two_way_ranging}
\end{figure}

Both platforms are based on the DW1000 chip, which provides the facilities for message time-stamping and precise control of message transmission times.
This enables a ranging method known as \ac{TWR}, where the \ac{ToF} between two nodes can be measured by exchanging packets and measuring their time of arrival, as illustrated in \cref{fig:two_way_ranging}.
The \ac{ToF} is then used for computing the distance that the \ac{RF} signal traveled to between the two nodes.

According to the DW1000's datasheet, a ranging accuracy of \SI{10}{\cm} can be achieved using \ac{UWB} and \ac{TWR}, based on the DW1000 clock precision.

\subsection{Platform}
\label{sec:platform}

The system is composed of two types of nodes.
So-called \emph{anchor} nodes are placed throughout the measurement environment at known locations.
So-called \emph{tag} nodes with unknown prior localization can then be localized by measuring their distance with the anchors via the \ac{TWR} process.

For performing our ranging measurements, the tag and anchors ran a modified version of Contiki~\cite{Dunkels2004,Corbalan2018}, an operating system designed for embedded sensor nodes.
A simple controller/agent architecture was developed, which ensures that only a single node can send a packet at a time, preventing packet collisions.
This also avoids more complex spectrum sharing mechanism.

For each ranging measurement between the tag and one of the anchors, the tag reports the resulting \ac{ToF}, the values of the various diagnostic registers from the DW1000, and the \ac{CIR} buffer to a central computer collecting all the data.
The \ac{CIR} contains information about the \ac{RF} signal properties of the underlying propagation paths in the measurement environment.

\label{sec:multilateration}

After the ranging process with all the anchors is finished, the multilateration phase is performed in order to compute the 2D localization of the tag.
The following optimization problem is solved:
\begin{equation}
	\label{eq:multilateration_opt}
	\min_{x_T,y_T} \sum_i \left((x_T - x_i)^2 + (y_T - y_i)^2 + \delta_z^2 - r_i^2\right)^2
\end{equation}
with $(x_T, y_T)$ the coordinates of the tag which need to be computed, $(x_i, y_i)$ the known coordinates of the anchor $i$, $\delta_z$ the known distance along the $z$ axis, and $r_i$ the distance between the tag and anchor $i$.
We use a standard least square method for solving \cref{eq:multilateration_opt}.


\section{Localization improvements}
\label{sec:localization_improvements}

The accuracy of the localization and multilateration approach presented earlier is highly dependent on the accuracy of the ranges and the measurement environment.
A first evaluation of the raw ranging values provided by the system showed an average absolute error of \SI{0.6}{\m} in our aircraft cabin, which lead to a poor localization accuracy, as shown later in \cref{sec:numerical_evaluation}.

Additionally, the DW1000 chip already provides mitigation techniques to overcome multipath effects in \ac{NLOS} environments \cite{DecawaveAPS006}.
Despite these, our evaluation still showed non-negligible ranging error.
We propose in this section different approaches for improving the overall accuracy of the system.

\subsection{Static offset}

For this first approach, we apply a static offset to each ranging measurement:
\begin{equation}
\mathit{corrected\,range} = \mathit{range} + o_i
\end{equation}
with $o_i$ the parameter of the model, fitted for each anchor $i$ in the system.
This is the simplest error correction model, which can compensate for static delays for the \ac{TWR} process which were not necessarily accurately calibrated.

As shown later, this simple model is able to correct the ranging measurement in \ac{LOS} conditions with good results.

\subsection{Linear regression}

An improvement of the previous model is to take into account the impact of the range on the error.
We model this effect using \acf{LR}:
\begin{equation}
	\mathit{corrected\,range} = \mathit{range} \cdot a_i + b_i
\end{equation}
with $a_i$ and $b_i$ parameters of the model, which are fitted for each anchor $i$ in the system.
This model accounts for a correlation between the measured ranges and the resulting ranging error.

\subsection{Neural network}
\label{sec:nn}

We propose in this section a \ac{NN}-based approach to predict various aspects about the localization.
We introduce here various extensions to our previous work \cite{Karadeniz2020}, with additional output types and a fully end-to-end training and prediction.

\begin{figure}[h!]
	\centering
	\includegraphics[width=.7\columnwidth]{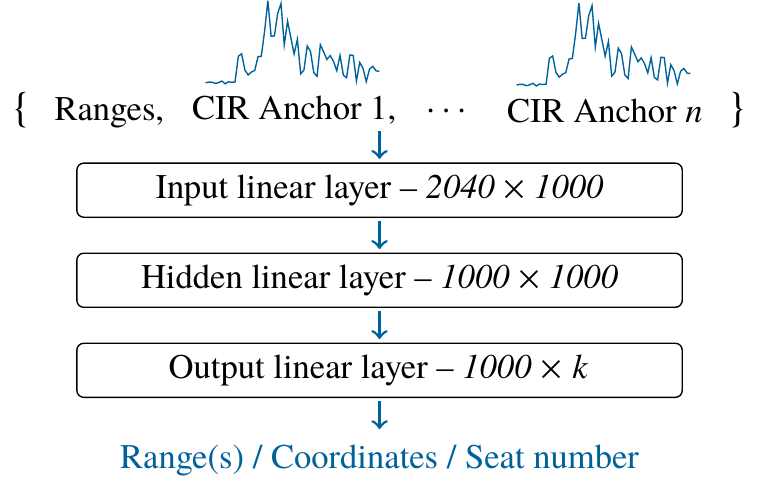}
	\caption{Architecture of the \ac{NN} and the size of the different layers. The size of the last layer depends on the type of output required.}
	\label{fig:nn_architecture}
\end{figure}

Our \ac{NN} architecture is illustrated in \cref{fig:nn_architecture}.
As input, a vector concatenating the measured ranges and \ac{CIR} data of one or all the anchors is used.
This vector is then processed by three fully-connected \ac{NN} layers with ReLU activation.
As output, the \ac{NN} predicts either the ranges, the coordinates of the tag, or the seat label associated to the tag's position as presented in \cref{tab:nn_variants}.

In total, we trained four different versions of the \ac{NN}, presented in \cref{tab:nn_variants}.
The first two versions (\emph{NN 1A} and \emph{NN range pred.}) still require the multilateration step presented in \cref{sec:multilateration,eq:multilateration_opt} in order to compute the 2D localization of the tag.
The last two versions follow an end-to-end approach and directly predict the tag's position, either as 2D coordinates or as label.
This means that the optimization step from \cref{eq:multilateration_opt} is not required for these versions in order to compute the 2D localization of the tag.

\begin{table}[h!]
	\caption{Variants of the neural network used in our evaluation}
	\label{tab:nn_variants}
	\centering
	\begin{tabular}{l|lll}
		\toprule
		\textbf{Label}           & \textbf{Inputs} & \textbf{Outputs}  & \textbf{Problem} \\ \midrule
		\textbf{NN 1A}           & 1 anchor        & 1 range           & Regression       \\
		\textbf{NN range pred.}  & All anchors     & All ranges        & Regression       \\
		\textbf{NN coord. pred.} & All anchors     & Coord. of the tag & Regression       \\
		\textbf{NN seat pred.}   & All anchors     & Seat label        & Classif.         \\ \bottomrule
	\end{tabular}
\end{table}

\begin{figure*}[!ht]
	\begin{subfigure}{0.32\textwidth}
		\includegraphics[width=\textwidth,clip,trim=0 520pt 0 450pt]{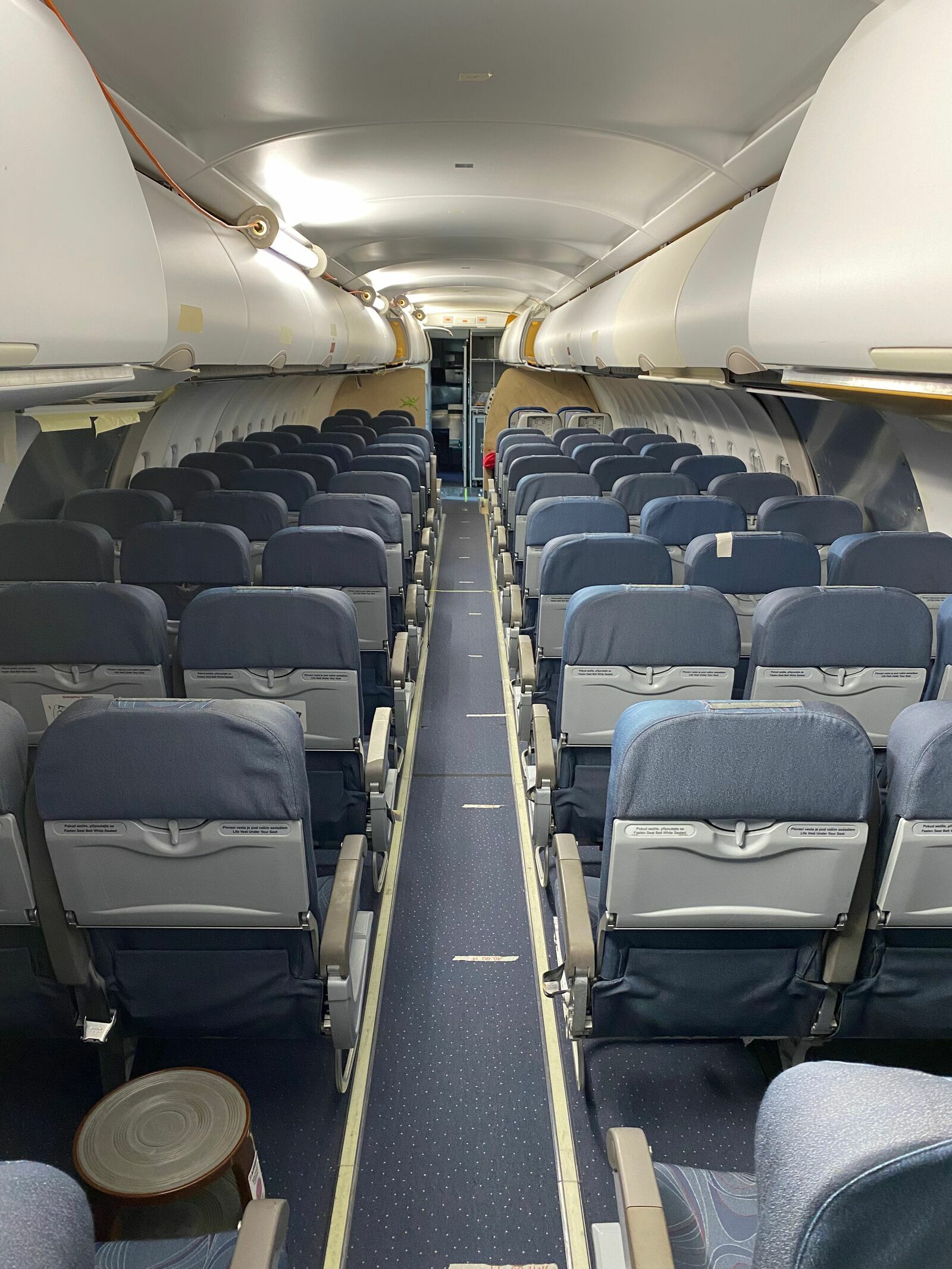}
		\caption{Aircraft cabin used for measurements}
		\label{fig:cabinphoto}
	\end{subfigure}
	\hfill
	\begin{subfigure}{0.32\textwidth}
		\includegraphics[width=\textwidth]{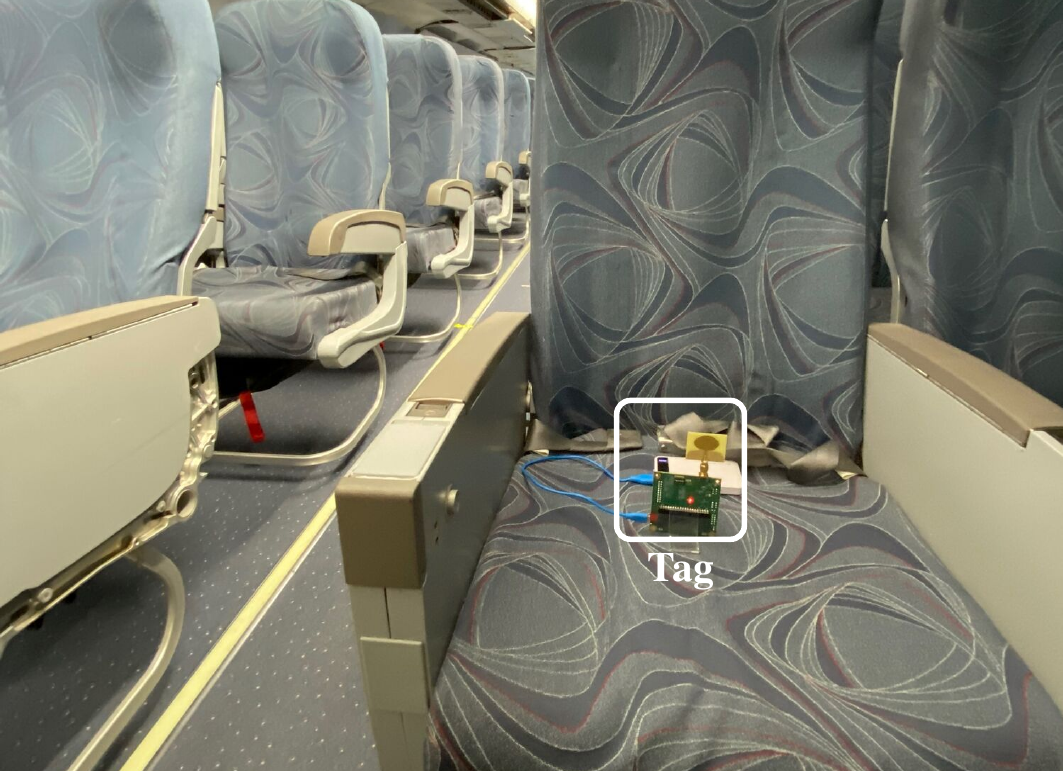}
		\caption{Example of tag positioned on seat}
		\label{fig:tag_on_seat}
	\end{subfigure}
	\hfill
	\begin{subfigure}{0.32\textwidth}
		\includegraphics[width=\textwidth]{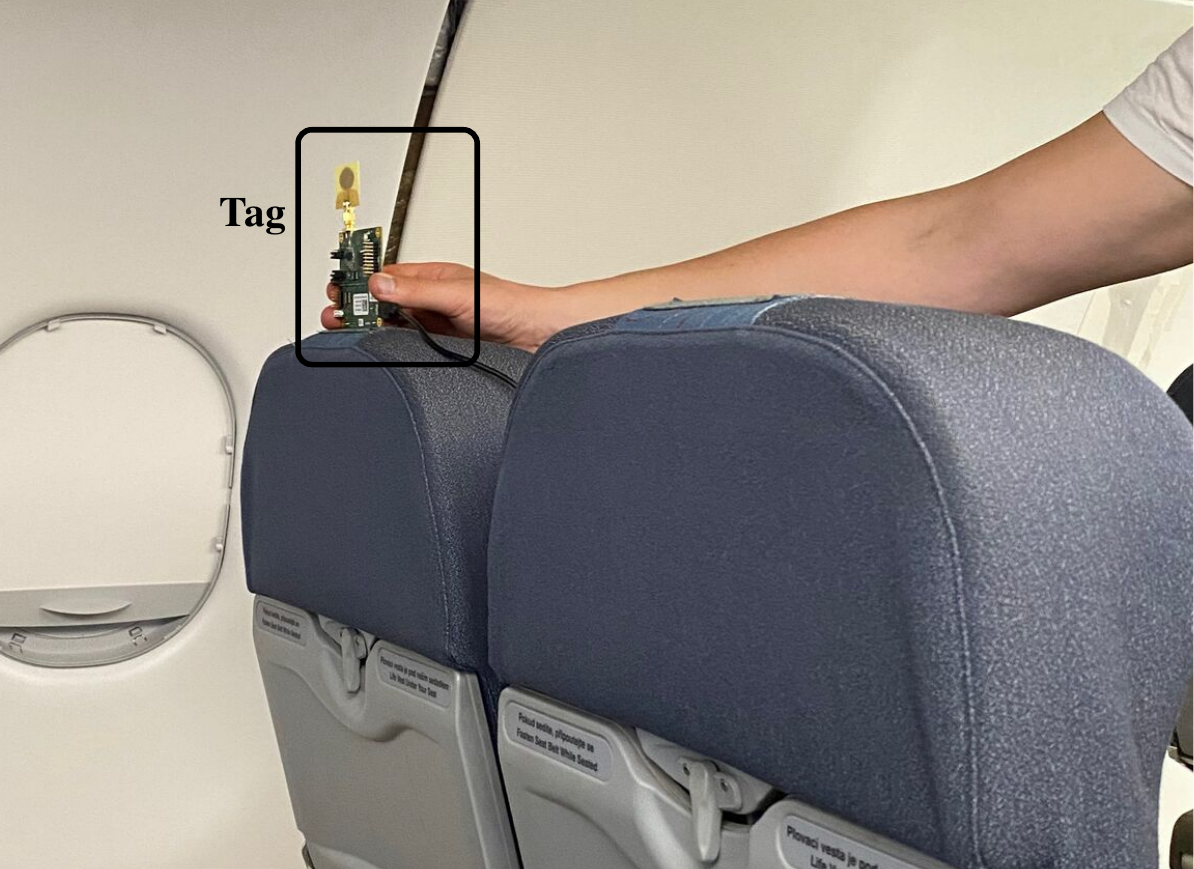}
		\caption{Example of tag positioned on headrest}
		\label{fig:tag_on_headrest}
	\end{subfigure}
	\caption{Measurement environment and tag positioning}
	\label{fig:measurement_environment}
\end{figure*}

The \acp{NN} were implemented and trained using PyTorch~\cite{Paszke2019}.
Hyper-parameter optimization was performed in order to finetune the parameters of the training process of the \acp{NN}.


\section{Numerical evaluation}
\label{sec:numerical_evaluation}

We describe in this section our measurement campaign made in real aircraft cabin, and the ranging and localization results which were measured.

\subsection{Measurement environments}

\begin{figure}[h!]
	\centering
	\includegraphics[width=\columnwidth,clip=true,trim=80pt 0 50pt 0]{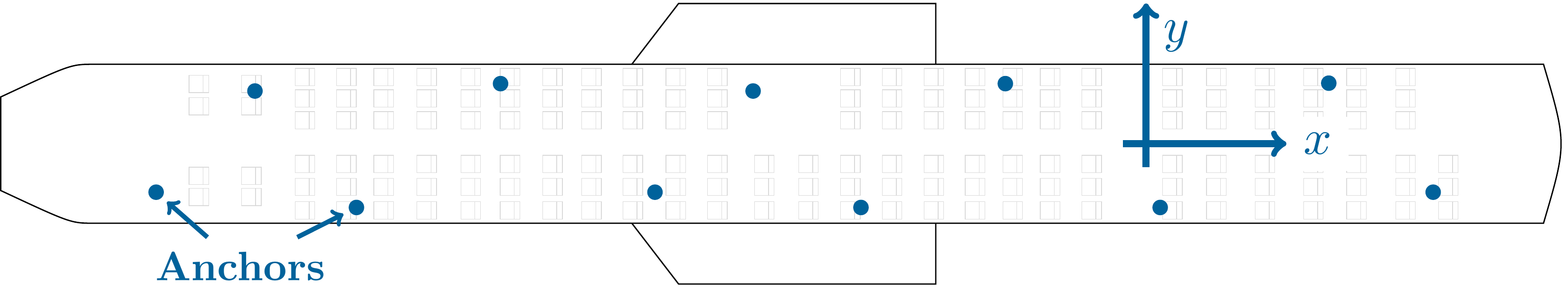}
	\caption{A321 cabin layout and anchor placements. The position of each seat was measured during our measurement campaign.}
	\label{fig:cabin_layout}
\end{figure}

Our measurements were performed in a real Airbus A321 fully furnished with 161 seats, specifically used for tests and measurements located at the Airbus manufacturing plant in Finkenwerder, Germany.
The interior cabin of this aircraft is illustrated in \cref{fig:cabinphoto}.
In total, 11 anchors were placed throughout the cabin according to \cref{fig:cabin_layout} using Qorvo's MDEK1001 development module.
In order to assess a potential industrial installation, the anchors were placed inside the luggage compartments as close to the cabin's hull as possible.

For our ranging measurements, Qorvo's EVK1000 evaluation kit was used as the tag to be localized.
All 161 seats in the cabin were measured, with two different positions: directly on the seat as illustrated on \cref{fig:tag_on_seat}, or on the headrest as illustrated on \cref{fig:tag_on_headrest}.
For each position, 10 different ranging measurements were performed, 7 of them used for training the various \acp{NN} presented in \cref{sec:nn} and 3 for the evaluation presented here.

As a comparison, we also performed measurements in two additional environments: outside and in an indoor office environment, both where no obstacle between the tag and anchors was present.
These additional environments will serve as references to estimate the impact of the cabin on the measurements.

The \ac{UWB} channel, communications, and ranging parameters used for our measurements are summarized in \cref{tab:uwb_parameters}.

\begin{table}[h!]
	\centering
	\caption{UWB channel parameters used for the measurement campaign}
	\label{tab:uwb_parameters}
	\begin{tabular}{ll}
		\toprule
		\textbf{Parameter} & \textbf{Value}        \\ \midrule
		UWB channel        & 4                     \\
		Center frequency   & \SI{3993.6}{\MHz}     \\
		Bandwidth          & \SI{1331.2}{\MHz}     \\
		Preamble length    & 128 symbols           \\
		Preamble code      & 17                    \\
		Data rate          & \SI{6.8}{Mbps}        \\
		Ranging method     & Single sided \ac{TWR} \\ \bottomrule
	\end{tabular}
\end{table}

\subsection{Impact of the environment on the properties of the physical signal}
\label{sec:impact_physics}

We first review the impact of the different measurement environments on the properties of the physical \ac{RF} signal.

\begin{figure}[h!]
	\centering
	\includegraphics[width=\columnwidth]{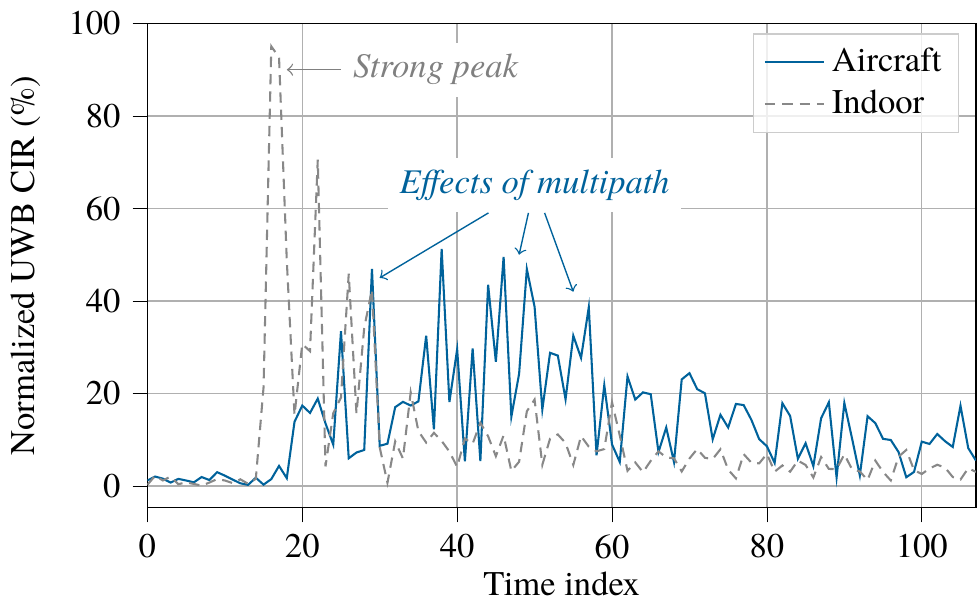}
	\caption{Example of \ac{CIR} data. The effects of multipath are clearly seen in the aircraft cabin compared to the indoor environment.}
	\label{fig:example_cir_data}
\end{figure}

Two samples of the \ac{CIR} data are presented in \cref{fig:example_cir_data}, one in the aircraft cabin, and the other in the indoor office environment, both for a distance between the nodes of approximately \SI{10}{\m}.
While both environments exhibit multipath effects, the magnitude of this effect is much larger in the aircraft environment.
We also notice a stronger fading of the \ac{RF} signal in the aircraft environment.

To numerically assess the fading of the \ac{RF} signal, we use the first path power level as our metric.
Based on Qorvo's documentation, we use the following estimation formula for computing this power level:
\begin{equation}
	10 \cdot \log_{10} \left( \frac{F_1^2 + F_2^2 + F_3^2}{N^2} \right) - A
\end{equation}
where $F_i$ corresponds to the first path amplitude point $i$ registers, and $A$ a constant dependent on the radio configuration.

\begin{figure}[h!]
	\centering
	\includegraphics[width=\columnwidth]{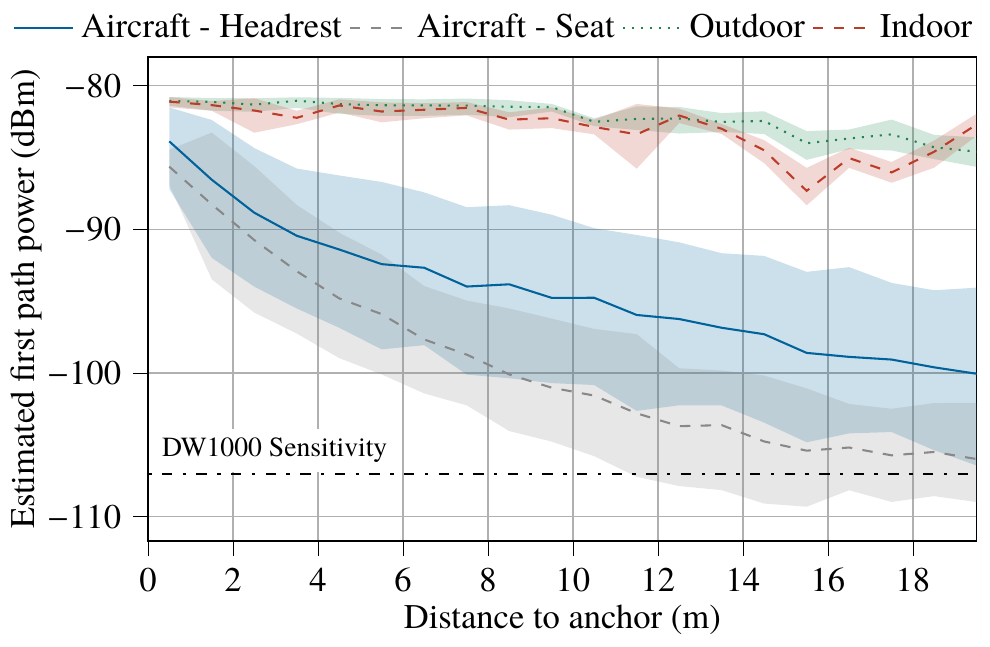}
	\caption{Impact of distance on estimated first path power level. Areas correspond to the \SI{25}{\percent} and \SI{75}{\percent} percentiles.}
	\label{fig:distance_vs_firstpathpowerlevel}
\end{figure}

The impact of the ranging distance on the first path power levels are presented in \cref{fig:distance_vs_firstpathpowerlevel}.
Compared with the outdoor and indoor environments, there is a much stronger attenuation on the received signal power in the aircraft, mainly due to the higher number of obstacles.

This strong attenuation had an impact on the connectivity, since ranging between the extreme front and back of the cabin was difficult due to the power levels below the DW1000 sensitivity margin.

\begin{figure}[!h]
	\includegraphics[width=\columnwidth]{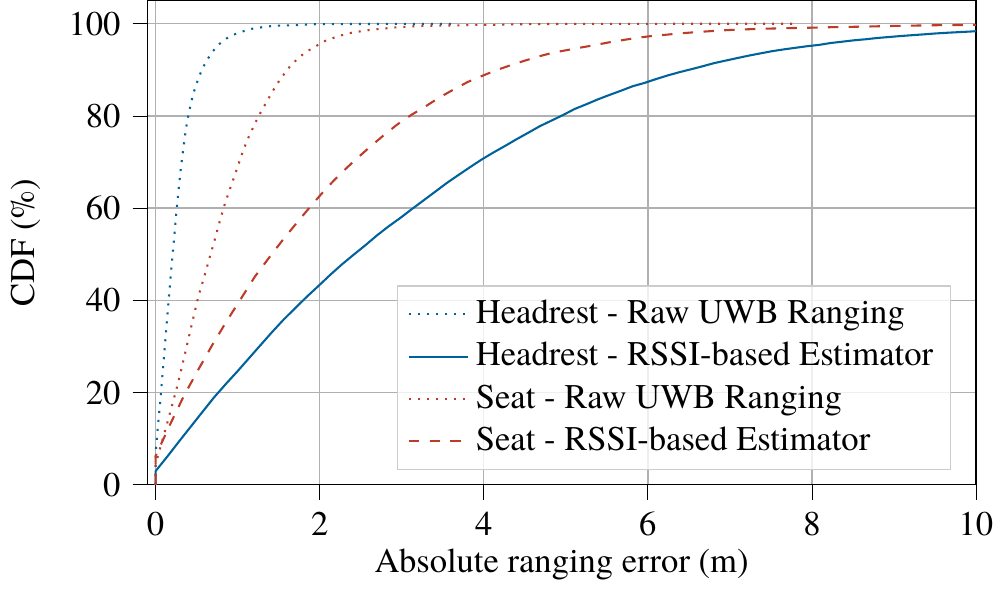}
	\caption{Absolute ranging error of an \ac{RSSI}-based distance estimator}
	\label{fig:ecdf_ranging_error_raw_and_rssibased}
\end{figure}

We note also in \cref{fig:distance_vs_firstpathpowerlevel} a correlation between the estimated first path power and the distance to anchor.
This relationship is often used in well-known localization methods such as Bluetooth-based or WiFi-based systems, which use \ac{RSSI} information for distance estimation.
To understand how such system would have performed in our measurement environment, we fitted a polynomial of degree 3 to the data and computed the resulting absolute ranging error of such a \ac{RSSI}-based system.
Results are presented in \cref{fig:ecdf_ranging_error_raw_and_rssibased}.
Compared to the non-corrected values provided by the \ac{UWB} ranging system, the \ac{RSSI}-based estimator provides worst results, with an average absolute error of \SI{2.52}{\m}.
This demonstrates why \ac{RSSI}-based systems are not sufficient for our use-cases and motivates our choice of a \ac{UWB}-based system.

Finally, we also numerically quantify the multipath effects visible on \cref{fig:example_cir_data} with the following metric:
\begin{equation}
\begin{array}{c}
\mathit{multipath} \\
\mathit{metric}
\end{array}
	 = \frac{1}{N - 1} \sum_{i=1}^N \left( 1 - \frac{\mathit{CIR}_i}{\max_k \mathit{CIR}_k} \right)
\end{equation}
A value close to 0 indicates a strong single path in the \ac{CIR} data, while a value close to 1 indicates the presence of multiple peaks with similar energy levels.
This is illustrated in \cref{fig:example_cir_data}, where the \ac{CIR} measured indoor would have a value close to 0 due to its strong peak, while the one measured in the cabin would have a larger value, closer to 1 due to the presence of multipaths.

\begin{figure}[h!]
	\centering
	\includegraphics[width=\columnwidth]{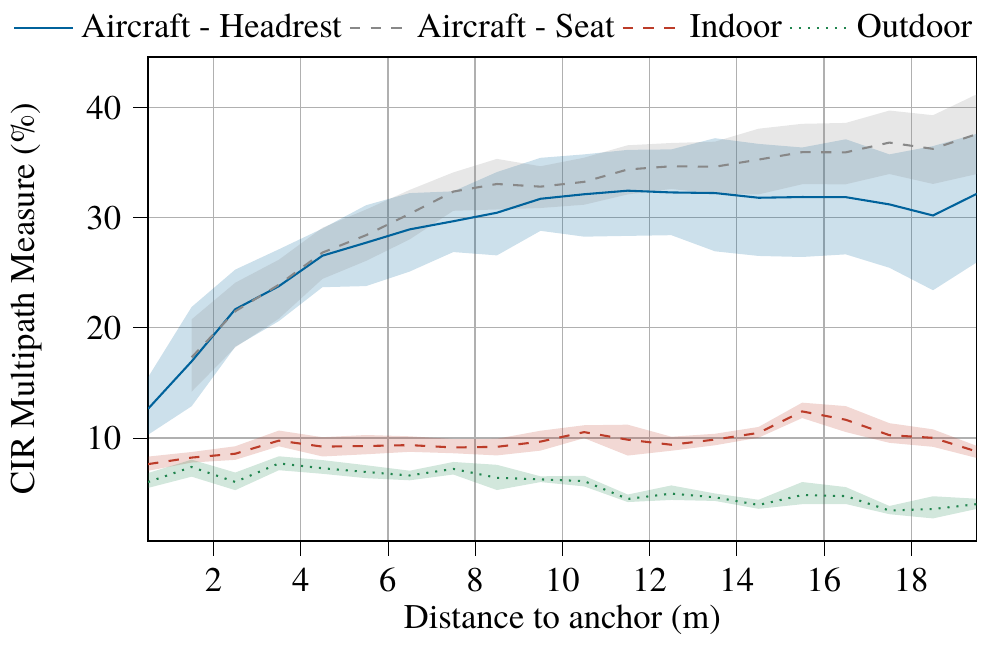}
	\caption{Impact of distance between anchor and tag on CIR multipath metric. Areas correspond to the \SI{25}{\percent} and \SI{75}{\percent} percentiles.}
	\label{fig:distance_vs_cir_multipath}
\end{figure}

Results with respect to the ranging distance are presented in \cref{fig:distance_vs_cir_multipath}.
As expected, there is a strong correlation between the ranging distance and the multipath metric in the aircraft cabin.
In the outdoor and indoor environments, the multipath metric remains close to constant.

Overall, \cref{fig:example_cir_data,fig:distance_vs_firstpathpowerlevel,fig:distance_vs_cir_multipath} illustrate the different challenges that an aircraft cabin poses and its impact on the properties of the \ac{RF} signals.
This will be illustrated in the next sections, where these effects result in large ranging errors.

\subsection{Ranging accuracy}
\label{sec:eval:ranging_accuracy}

We evaluate in this section the ranging accuracy, i.e. the accuracy of the individual distance measurements between tag and anchors.
We use the absolute difference between measured and true range as our metric:
\begin{equation}
	\mathit{ranging\,error} = |\mathit{range} - \mathit{range}_\mathit{true}|
\end{equation}

\Cref{tab:ranging_acc_summary} provides a summary of the impact of the different correction methods presented in \cref{sec:localization_improvements} on the absolute ranging error in the both the outdoor and the cabin environment.
In the outdoor environment, we achieve decimeter level accuracy with the offset and \ac{LR} corrections, with an average of \SI{4}{\cm} which matches the advertised accuracy of the DW1000 system.

\begin{figure}[h!]
	\centering
	\begin{subfigure}{\columnwidth}
		\includegraphics[width=\columnwidth]{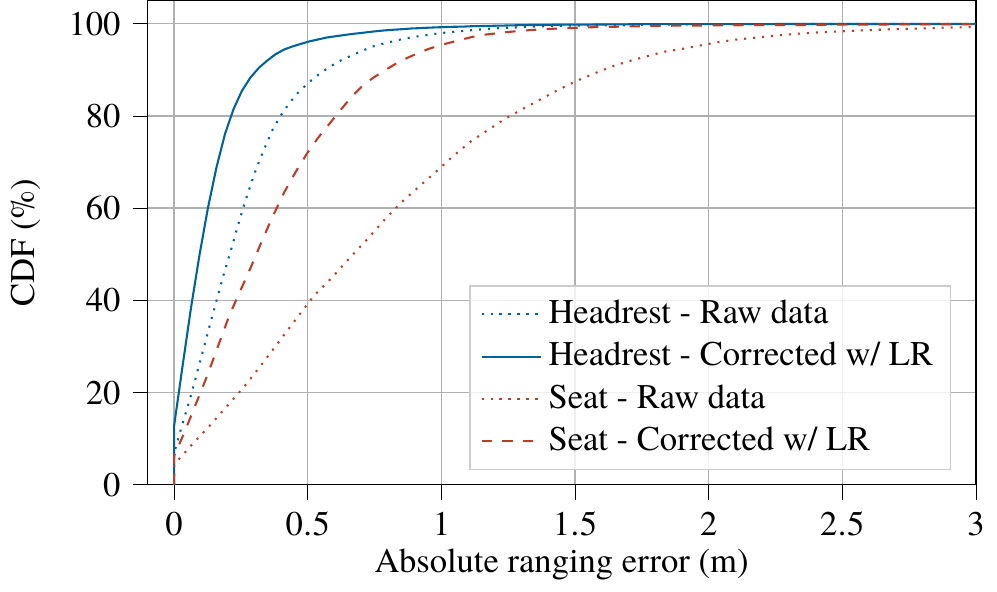}
		\caption{Absolute ranging error before and after correction with \ac{LR}}
		\label{fig:ecdf_ranging_error_raw_and_lrcorrected}
	\end{subfigure}
	\vspace{10pt}
	
	\begin{subfigure}{\columnwidth}
		\includegraphics[width=\columnwidth]{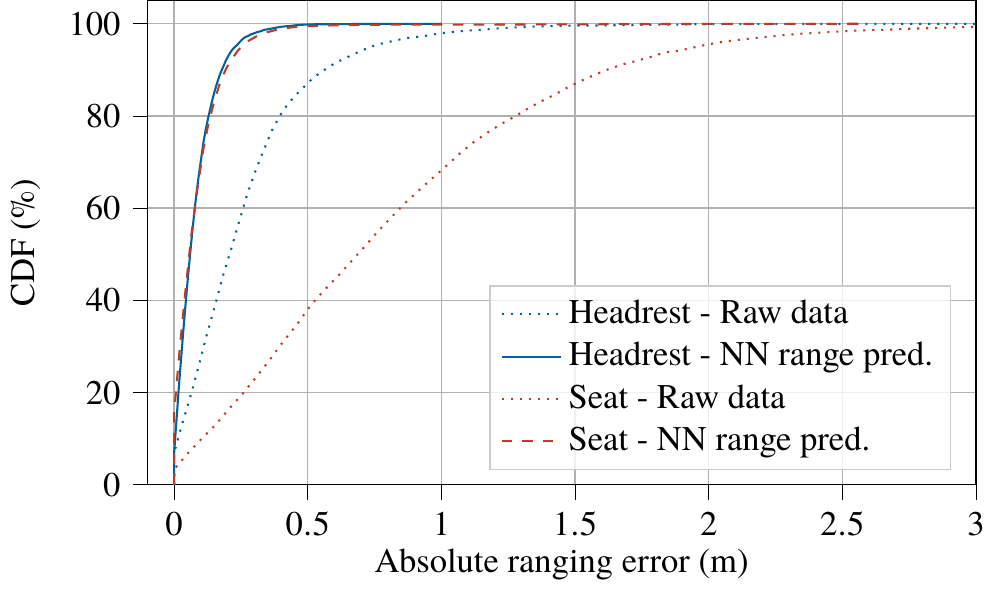}
		\caption{Absolute ranging error before and after correction with \ac{NN}}
		\label{fig:ecdf_ranging_error_raw_and_nncorrected}
	\end{subfigure}
	\caption{Absolute ranging errors with \ac{LR} and \ac{NN}}
\end{figure}

In the aircraft environment, we notice a larger ranging error of \SI{30}{\cm} in average even after offset and \ac{LR} correction.
This is especially visible at the tail of the statistical distribution as shown by the \SI{95}{\percent} quantile and later in \Cref{fig:ecdf_ranging_error_raw_and_lrcorrected,fig:ecdf_ranging_error_raw_and_nncorrected}.

\begin{table}[h!]
	\centering
	\caption{Summary of the absolute ranging error before and after correction methods in the aircraft and outdoor environments}
	\label{tab:ranging_acc_summary}
	\begin{tabular}{ll|ccc}
		\toprule
		\textbf{Env.}             & \textbf{Method} &       \textbf{Mean}       &      \textbf{Median}      & \textbf{Q-\SI{95}{\percent}} \\ \midrule
		\multirow{2}{*}{Outdoor}  & Raw data        &      \SI{0.389}{\m}       &      \SI{0.430}{\m}       &        \SI{0.499}{\m}        \\
		                          & Corr. w/ Offset &      \SI{0.077}{\m}       &      \SI{0.045}{\m}       &        \SI{0.275}{\m}        \\
		                          & Corr. w/ LR     & {\bfseries\SI{0.040}{\m}} & {\bfseries\SI{0.033}{\m}} &  {\bfseries\SI{0.103}{\m}}   \\ \midrule
		\multirow{4}{*}{Aircraft} & RSSI estimator  &      \SI{2.520}{\m}       &      \SI{1.944}{\m}       &        \SI{6.835}{\m}        \\
		                          & Raw data        &      \SI{0.604}{\m}       &      \SI{0.411}{\m}       &        \SI{1.728}{\m}        \\
		                          & Corr. w/ Offset &      \SI{0.335}{\m}       &      \SI{0.223}{\m}       &        \SI{0.968}{\m}        \\
		                          & Corr. w/ LR     &      \SI{0.307}{\m}       &      \SI{0.211}{\m}       &        \SI{0.892}{\m}        \\
		                          & NN 1A           &      \SI{0.160}{\m}       &      \SI{0.113}{\m}       &        \SI{0.486}{\m}        \\
		                          & NN range pred.  & {\bfseries\SI{0.044}{\m}} & {\bfseries\SI{0.020}{\m}} &  {\bfseries\SI{0.169}{\m}}   \\ \bottomrule
	\end{tabular}
\end{table}

The \ac{NN}-based error correction brings a large benefit, almost matching the accuracy measured in the outdoor environment.
Despite larger errors, the variant of the \ac{NN} only using the input of one anchor (labeled \emph{NN 1A} in \cref{tab:ranging_acc_summary}) still brings a benefit compared with the \ac{LR} correction.

\Cref{fig:ecdf_ranging_error_raw_and_lrcorrected,fig:ecdf_ranging_error_raw_and_nncorrected} present additional details about the impact of the \ac{LR} and \ac{NN} corrections on the ranging accuracy.
As expected from the conclusions from \cref{sec:impact_physics}, the ranging error when the tag is placed on the seat is larger compared to the headrest.

We note on \cref{fig:ecdf_ranging_error_raw_and_nncorrected} that this difference between the two positions is less relevant when using the \ac{NN}, as there is strong overlap between the two curves.

\subsection{Localization accuracy}
\label{sec:eval:localization_accuracy}

We evaluate in this section the localization accuracy, i.e. the accuracy of the computed 2D coordinates based on the ranging measurements.
We use the Euclidean distance to the true coordinates as our metric:
\begin{equation}
	\mathit{localization\,error} = \sqrt{(x_T - x_\mathit{true})^2 + (y_T - y_\mathit{true})^2}
\end{equation}
with $(x_T, y_T)$ the computed coordinates of the tag after multilateration or the predicted coordinates from the \ac{NN}.

\begin{figure}[h!]	
	\centering
	\begin{subfigure}{\columnwidth}
		\includegraphics[width=\columnwidth]{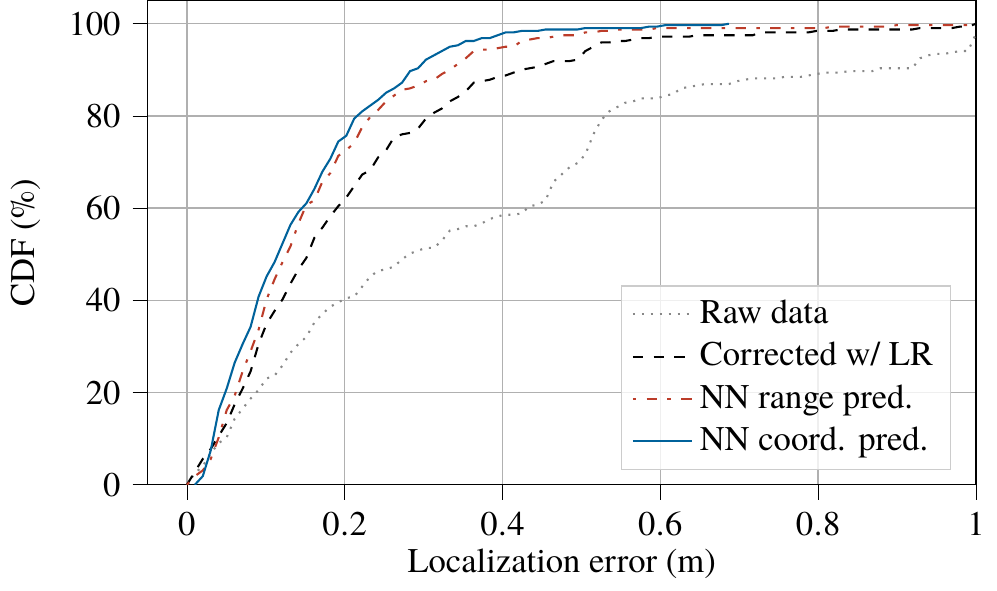}
		\caption{Tag placed at the headrest positions}
		\label{fig:ecdf_localization_headrest_error_raw_and_nncorrected}
	\end{subfigure}
	\vspace{10pt}

	\begin{subfigure}{\columnwidth}
		\includegraphics[width=\columnwidth]{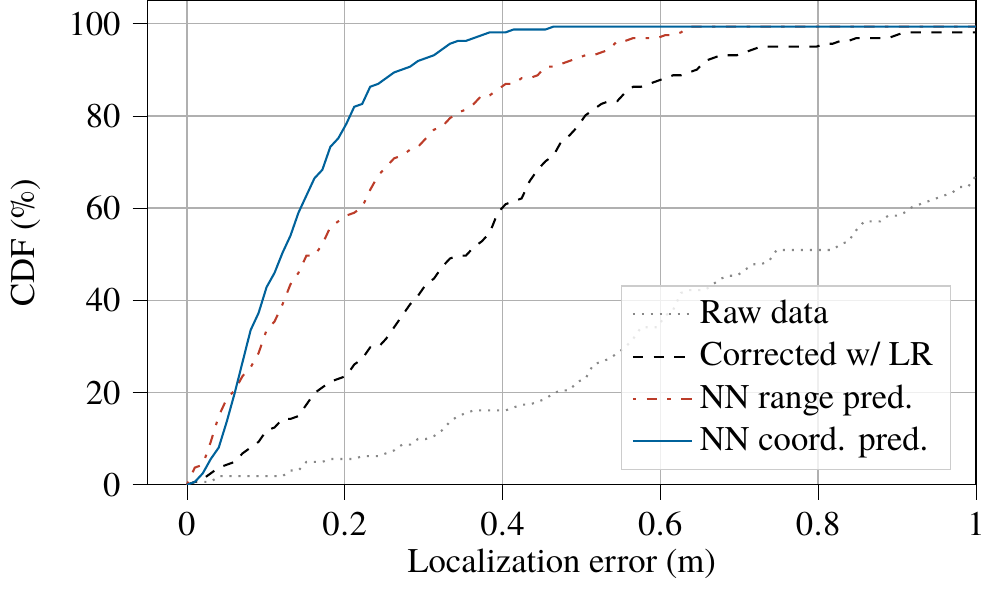}
		\caption{Tag placed at the seat positions}
		\label{fig:ecdf_localization_seat_error_raw_and_nncorrected}
	\end{subfigure}
	\caption{Impact of the different correction methods on the localization error of our system}
\end{figure}

\Cref{tab:summary_loc_accuracy} provides a summary of the impact of the corrections from \cref{sec:localization_improvements} on the localization error.
As for the previous results, the larger ranging errors also lead to larger localization errors when the tag is placed on the seat compared with when it is placed on the headrest.
\Cref{fig:ecdf_localization_headrest_error_raw_and_nncorrected,fig:ecdf_localization_seat_error_raw_and_nncorrected} present additional details about the impact of the \ac{LR} and \ac{NN} corrections on the localization accuracy.

\begin{table}[h!]
	\caption{Summary of localization accuracy in the aircraft cabin}
	\label{tab:summary_loc_accuracy}
	\centering
\begin{tabular}{l|cc|cc}
	\toprule
	                &        \multicolumn{2}{c|}{\textbf{Tag on seat}}         &       \multicolumn{2}{c}{\textbf{Tag on headrest}}       \\
	\textbf{Method} &       \textbf{Mean}       & \textbf{Q-\SI{90}{\percent}} &       \textbf{Mean}       & \textbf{Q-\SI{90}{\percent}} \\ \midrule
	Raw data        &      \SI{0.923}{\m}       &        \SI{1.883}{\m}        &      \SI{0.354}{\m}       &        \SI{0.871}{\m}        \\
	Corrected w/ LR &      \SI{0.379}{\m}       &        \SI{0.639}{\m}        &      \SI{0.199}{\m}       &        \SI{0.418}{\m}        \\
	NN range pred.  &      \SI{0.208}{\m}       &        \SI{0.447}{\m}        &      \SI{0.161}{\m}       &        \SI{0.332}{\m}        \\
	NN coord. pred. & {\bfseries\SI{0.167}{\m}} &  {\bfseries\SI{0.268}{\m}}   & {\bfseries\SI{0.142}{\m}} &  {\bfseries\SI{0.288}{\m}}   \\ \bottomrule
\end{tabular}
\end{table}

Regarding the \ac{NN} variants, the \ac{NN} directly predicting the coordinates appears to result in a better accuracy compared with the \ac{NN} predicting the only ranges.
This is explained by the additional multilateration process required for the range variant, since small errors on the ranging can propagate in the multilateration computation, leading to larger errors.
This illustrates that an end-to-end training approach is more effective for the localization process.

\subsection{Seat assignment}
\label{sec:eval:seat_assignment}

Finally, we evaluate in this section the use-case where the tag has to be assigned to a seat based on its localization.
This is performed by assigning it to the closest known seat position given the computed coordinates.
This is formalized as:
\begin{equation}
\begin{array}{c}
\mathit{assigned} \\
\mathit{seat}
\end{array} = \underset{\mathit{seat}}{\arg\min} \sqrt{(x_T - x_\mathit{seat})^2 + (y_T - y_\mathit{seat})^2}
\end{equation}
with $(x_\mathit{seat}, y_\mathit{seat})$ the known locations of the different seats in the aircraft.

According to the cabin geometry presented in \cref{fig:cabin_layout}, a seat assignment accuracy of \SI{100}{\percent} would require a localization system with a maximum error of \SI{0.226}{\m}.
Following the results from \cref{tab:summary_loc_accuracy}, our localization system is close to reaching this required accuracy.

Accuracy results of the different methods are presented in \cref{tab:seat_assign_acc}.
Since seat labels represent 2D coordinates (eg. seat \emph{5A} implies row \emph{5} and column \emph{A}), we split the accuracy in \cref{tab:seat_assign_acc} according to the $x$ axis (i.e. seat row axis) and the $y$ axis (i.e. seat letter axis) illustrated in \cref{fig:cabin_layout}.

For most methods, we notice a large difference between the two axes, where the accuracy along the $x$ axis is better than on the $y$ axis.
This is explained by the geometry of the aircraft cabin and the anchor placements within the cabin, making it more sensitive to small errors along the $y$ axis.

\begin{table}[h!]
	\centering
	\caption{Seat assignment accuracy of the different methods}
	\label{tab:seat_assign_acc}
\begin{tabular}{c|l|c|cc}
	\toprule
	                   \textbf{Tag}                    &                 &         \textbf{Seat}          &      \textbf{{\it x} Axis}      &     \textbf{{\it y} Axis}      \\
	                  \textbf{Pos.}                    & \textbf{Method} &       \textbf{Accuracy}        &        \textbf{Accuracy}        &       \textbf{Accuracy}        \\ \midrule
	\multirow{5}{*}{\rotatebox{90}{\textbf{Headrest}}} & Raw data        &      \SI{52.0}{\percent}       &       \SI{99.1}{\percent}       &      \SI{52.6}{\percent}       \\
	                                                   & Corrected w/ LR &      \SI{74.8}{\percent}       &       \SI{99.7}{\percent}       &      \SI{75.1}{\percent}       \\
	                                                   & NN range pred.  &      \SI{83.8}{\percent}       & {\bfseries\SI{100.0}{\percent}} &      \SI{83.8}{\percent}       \\
	                                                   & NN coord. pred. &      \SI{87.5}{\percent}       & {\bfseries\SI{100.0}{\percent}} &      \SI{87.5}{\percent}       \\
	                                                   & NN direct pred. & {\bfseries\SI{97.2}{\percent}} &       \SI{99.7}{\percent}       & {\bfseries\SI{99.1}{\percent}} \\ \midrule
	  \multirow{5}{*}{\rotatebox{90}{\textbf{Seat}}}   & Raw data        &      \SI{18.6}{\percent}       &       \SI{60.2}{\percent}       &      \SI{36.6}{\percent}       \\
	                                                   & Corrected w/ LR &      \SI{60.9}{\percent}       & {\bfseries\SI{100.0}{\percent}} &      \SI{64.6}{\percent}       \\
	                                                   & NN range pred.  &      \SI{75.2}{\percent}       & {\bfseries\SI{100.0}{\percent}} &      \SI{77.6}{\percent}       \\
	                                                   & NN coord pred.  &      \SI{91.9}{\percent}       &       \SI{99.4}{\percent}       &      \SI{96.9}{\percent}       \\
	                                                   & NN direct pred. & {\bfseries\SI{98.8}{\percent}} & {\bfseries\SI{100.0}{\percent}} & {\bfseries\SI{98.8}{\percent}} \\ \bottomrule
\end{tabular}
\end{table}

The results from \cref{tab:seat_assign_acc} match the ones from \cref{sec:eval:localization_accuracy} and the requirement of \SI{0.226}{\m} localization error.
As previously, the \ac{NN} directly predicting the seat label also achieves the best results, with an accuracy better than \SI{97}{\percent} in both use-cases.
This illustrates again the impact of an end-to-end approach for training a \ac{NN}.

\subsection{Localization speed}

We detail in this section the time required for the localization process.
As illustrated in \cref{fig:two_way_ranging}, our system requires the exchange of two messages with each anchor, which require less than \SI{1}{\ms} per anchor.
The collected ranges and \ac{CIR} data are then processed, either by the multilateration process by solving \cref{eq:multilateration_opt} or by executing the \ac{NN}.

\Cref{tab:nn_speed} illustrates the processing time required for the \ac{NN} on three different platforms: a server with an Intel CPU and a Nvidia GPU representative of servers used in datacenters, and the last two generations of the Raspberry Pi.
Two different implementations of the \ac{NN} are presented here: PyTorch v1.10~\cite{Paszke2019}, and Genann~v1.0\footnote{\texttt{https://github.com/codeplea/genann}}.
While PyTorch is designed for performance and can also take advantage of GPU acceleration, we also selected Genann as a reference worst-case implementation since it does not rely on any hardware acceleration such as GPU, CPU extensions such as \ac{SIMD} instructions, or parallelization.

\begin{table}[!h]
	\centering
	\caption{Execution time of the \ac{NN} inference on different hardware platforms and \ac{NN} frameworks}
	\label{tab:nn_speed}
	\begin{tabular}{llr}
		\toprule
		\textbf{Hardware platform}                 & \textbf{Implem.} & \textbf{Avg. time} \\ \midrule
		Xeon 5120 at \SI{2.20}{GHz} + RTX 2080 GPU & PyTorch          &      \SI{0.2}{\ms} \\
		Xeon 5120 at \SI{2.20}{GHz}                & Genann           &      \SI{2.6}{\ms} \\
		Raspberry Pi 4  Model B+ Rev 1.4           & Genann           &      \SI{5.4}{\ms} \\
		Raspberry Pi 3  Model B+ Rev 1.3           & Genann           &     \SI{31.9}{\ms} \\ \bottomrule
	\end{tabular}
\end{table}

Overall, the \ac{NN} is fast even on low-cost hardware platforms, mainly due to the small sizes of the layers and a single hidden layer.
These values illustrate that the \ac{NN} inference is not a bottleneck in the localization system.


\section{Improvement options}
\label{sec:discussion}

With our measurement campaign and our evaluation in \cref{sec:numerical_evaluation}, we showed that our \ac{UWB}-based localization system is able to achieve an accuracy sufficient for assigning about \SI{97}{\percent} of the seats in a real aircraft cabin.

We discuss in this section various ways to further improve this accuracy, especially in conditions where the end-to-end training a \ac{NN} might not be possible.
We use Monte Carlo simulations to evaluate alternate configurations and their impact on the localization accuracy.

\subsection{Increased anchors count}

One obvious approach for improving the results is to add additional anchors throughout the cabin.
Since the localization error is correlated with the distance between anchor and tag, having more anchors would increasing the number of anchors closer to the tag, thus reducing the overall ranging error.

To evaluate this option, we performed Monte Carlo simulations where we simulated additional (virtual) anchors.
The additional anchors were randomly placed throughout the cabin.
The ranging error from the original anchors was statistically fitted to a Johnson's S\textsubscript{$U$}-distribution, which was then used for generating the ranging errors for the additional anchors:
\begin{equation}
	\mathit{simulated\,range} = \mathit{true\,range} + \mathit{noise}
\end{equation}
This specific distribution was chosen because it resulted in the best fit among a set of 50 other statistical distributions.

Based on these additional ranges, we performed the \ac{LR} correction and multilateration operations to compute the coordinates of the tag.
Multiple runs of this process were performed and the best localization errors were kept.

\begin{figure}[h!]
	\centering
	\includegraphics[width=\columnwidth]{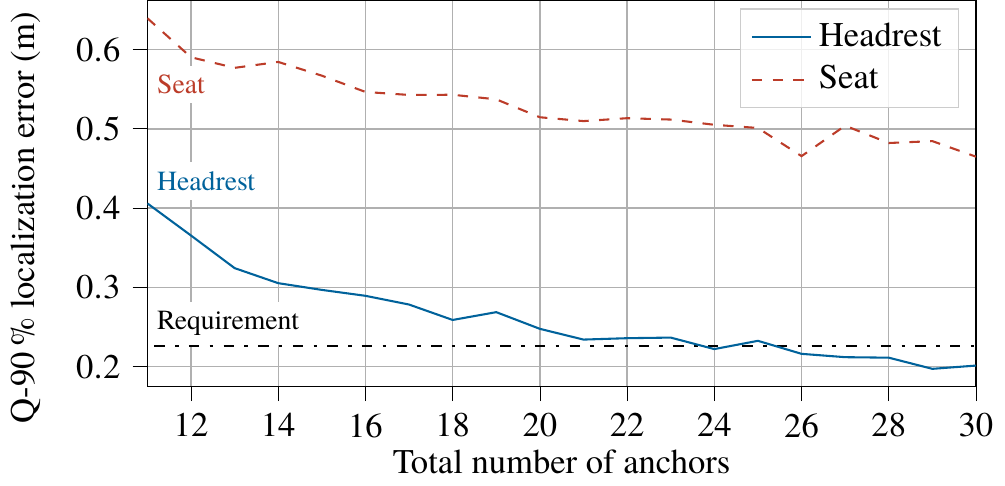}
	\caption{Impact of increased number of anchors on localization error via Monte-Carlo simulations}
	\label{fig:simulation_more_anchors}
\end{figure}

Results are presented in \cref{fig:simulation_more_anchors}.
Compared with our original measurements with 11 anchors, the additional anchors indeed result in a better localization accuracy.
By doubling the number of anchors, it would be possible to match the \SI{0.226}{\m} requirement for the headrest position.
For the seat position, the increased number of anchors is not sufficient for reaching the required localization accuracy.

\subsection{Better ranging accuracy}

Another approach for improving the results would be to reduce the ranging error of the system.
This may be achieved with alternate anchor placements, alternate antennas, or by improving the properties of the \ac{UWB} ranging system.
To simulate such approach, the ranging error of each measurement was scaled according to the formula
\begin{equation}
\mathit{simulated\,error} = \mathit{measured\,error} \cdot \alpha
\end{equation}
with $\alpha$ a scaling factor in the $(0, 1]$ interval.
As previously, we then performed the \ac{LR} correction and multilateration operations to compute the coordinates of the tag based on the ranges with scaled error.

\begin{figure}[h!]
	\centering
	\includegraphics[width=\columnwidth]{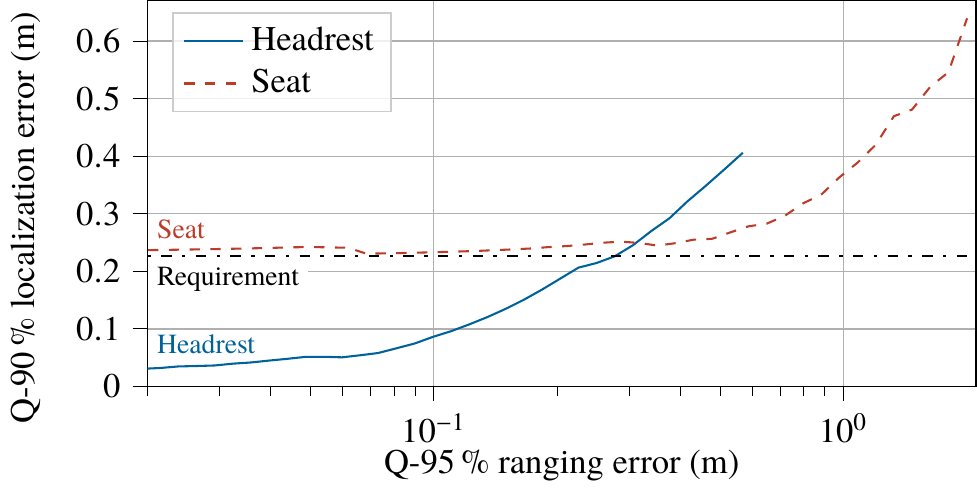}
	\caption{Impact of improved ranging accuracy on localization error via Monte-Carlo simulations}
	\label{fig:simulation_better_ranging_accuracy}
\end{figure}

Results are presented in \cref{fig:simulation_better_ranging_accuracy}.
Scaling the errors indeed lead to a fulfilled requirement for the headrest position.
For the seat position, we note that a better localization accuracy would be required.

\subsection{Other approaches}

We showed in \cref{sec:eval:seat_assignment} that the assignment accuracy is sensitive along the $y$ axis of the cabin.
Additional anchors along this axis should help improving this accuracy.
For our aircraft use-case, anchors could be placed along the wings of the aircraft, yet additional measurements would be required to validate if ranging would be possible in such conditions.


\acresetall
\section{Conclusion}
\label{sec:conclusion}

We presented in this work an \ac{UWB} based localization system targeting use-cases in aircraft cabins.
Our main contribution is a measurement campaign performed onboard a real Airbus A321 aircraft, as well as various methods for correcting ranging errors based on \ac{LR} and \acp{NN}.

Our measurements results confirm our previous findings from \cite{Karadeniz2020}.
While our previous measurements were performed in a mockup of a small cabin section, the results presented here illustrate that \ac{UWB} is indeed a valid candidate for an onboard aircraft localization system.
With our end-to-end \ac{NN} approach, we were able to localize a tag with an average localization error of about \SI{16}{\cm} and assign it to a seat with an accuracy of \SI{97}{\percent}.
We also show that compared against a simpler system using only \ac{RSSI} -- such as Bluetooth-based localization systems -- \ac{UWB} results in more accurate localization.
Our measurements also confirm that an aircraft cabin is a challenging environment for localization, due to the presence of many obstacles and propensity for multipath effects.

Overall, we demonstrate the viability of the \ac{UWB} technology for localization for an application in the aircraft industry.
Via Monte Carlo simulations, we also explored various methods for increasing the accuracy when using the \ac{LR} approach.


\section*{Acknowledgment}

The authors would like to thank Thomas Multerer and Thomas Meyerhoff for their contributions in an early stage of this research, and Susan Buchholz and her team at Rapid Architecture Lab for their assistance with the measurements.

\bibliography{IEEEabrv,biblio}
\bibliographystyle{IEEEtranN}

\end{document}